# Ultrafast switching dynamics of the ferroelectric order in stacking-engineered ferroelectrics

Ri He[1#], Bingwen Zhang[1,2#], Hua Wang[3], Lei Li[4], Tang Ping[5], Gerrit Bauer[5,6], Zhicheng Zhong[1,7*]

[1]Key Laboratory of Magnetic Materials Devices & Zhejiang Province Key Laboratory of Magnetic Materials and Application Technology, Ningbo Institute of Materials Technology and Engineering, Chinese Academy of Sciences, Ningbo 315201, China

[2]Fujian Key Laboratory of Functional Marine Sensing Materials, Center for Advanced Marine Materials and Smart Sensors, College of Material and Chemical Engineering, Minjiang University, Fuzhou 350108, China

[3]School of Micro-Nano Electronics, Hangzhou Global Scientific and Technological Innovation Center, Zhejiang University, Hangzhou 310027, China

[4]Frontiers Science Center for Flexible Electronics, Xi'an Institute of Flexible Electronics (IFE) and Xi'an Institute of Biomedical Materials & Engineering, Northwestern Polytechnical University, Xi'an 710072, China

[5]WPI-AIMR, Tohoku University, 2-1-1 Katahira, Sendai 980-8577, Japan

[6]Kavli Institute for Theoretical Sciences, University of the Chinese Academy of Sciences, Beijing 10090, China

[7]China Center of Materials Science and Optoelectronics Engineering, University of Chinese Academy of Sciences, Beijing 100049, China

---

[#] These authors contribute equally to this work.

[*] zhong@nimte.ac.cn

## *Abstract*

The recently discovered ferroelectricity of van der Waals bilayers offers an unconventional route to improve the performance of devices. Key parameters such as switching field and speed depend on the static and dynamic properties of domain walls (DWs). Here we theoretically explore the properties of textures in stacking-engineered ferroelectrics from first principles. Employing a machine-learning potential model, we present results of large-scale atomistic simulations of stacking DWs and Moiré structure of boron nitride bilayers. We predict that the competition between the switching barrier of stable ferroelectric states and the in-plane lattice distortion leads to a DW width of the order of ten nanometers. DWs motion reduces the critical ferroelectric switching field of a monodomain by two orders of magnitude, while high domain-wall velocities allow domain switching on a picosecond-timescale. The superior performance compared to conventional ferroelectrics (or ferromagnets) may enable ultrafast and power-saving non-volatile memories. By twisting the bilayer into a stacking Moiré structure, the ferroelectric transforms into a super-paraelectric since DWs move under ultralow electric fields.

Ferroelectric materials display a switchable spontaneous electric polarization, that usually emerges from ion displacements. By contrast, the spontaneous polarization in recently discovered stacking-engineered ferroelectrics of two van der Waals monolayers originates from the interlayer charge transfer. It can be controlled by interlayer sliding without any vertical ion displacements [1,2] and leads to ultralow ferroelectric switching barriers and high thermal stability [3-6]. The research of stacking-engineered ferroelectricity is still in its infancy since many fundamental and practical issues associated with the emerging ferroelectricity remain to be settled [7,8]. Here, we focus on two fundamental observations: (i) the experimentally observed critical switching field is an order magnitude smaller than the theoretical value [1,6,7], and (ii) the Moiré structure caused by twisting the bilayers significantly reduces the coercive field [3,9].

An in-depth understanding of the atomistic characteristics of ferroelectric DWs in stacked van der Waals bilayers [10,11] is crucial in solving the above two questions. A ferroelectric DW is a topological defect that affects various properties such as electric conductivity [12,13], dielectric permittivity [14], multiferroic phenomena [15,16], and the photovoltaic effect [17,18]. Moreover, DW motion driven by an external field facilitates ferroelectric switching [19-21]. However, the rotation of the polarization in Bloch and Néel DWs of conventional ferroelectrics costs significant elastic energy [10,11,20,21]. On the other hand, DWs in stacking ferroelectrics with weak van der Waals interactions remain yet unexplored, experimentally and theoretically [8]. Indeed, the complexity of DWs in stacking-engineered ferroelectric far exceeds the computing capacity of traditional theoretical first-principles methods such as density functional theory (DFT) with a computation time that scales polynomially with the unit cell size.

In this Letter, we report a machine-learning-based Deep Potential (DP) model using training data from DFT calculations that allows us to systematically and directly study large-scale DWs and Moiré structures in hexagonal boron nitride (*h*-BN) bilayers with DFT accuracy. We find that the 0° and 90° DWs expose Bloch- and Néel-type textures, respectively. Their widths turn out to be 10~40 nm, while the critical switching field facilitated by DW motion is two orders of magnitude smaller than that for monodomain reversal. Ferroelectric skyrmions can be controlled by perpendicular gate electric fields at room temperature, enabling ultra-dense information storage devices [22] without need for the weak relativistic spin-orbit couplings that govern magnetic skyrmions. Finally, we demonstrate that a twisted Moiré structure is a super-paraelectric because the ferroelectric order is reversibly broken by domain wall motion already at ultralow electric fields.

***Stacking modes and thermal stability.*** −The *h*-BN stacking configurations can be divided into two categories, i.e. with parallel and antiparallel twists [3,7,23-25]. Bulk *h*-BN crystals favor the centrosymmetric antiparallel AA′ structure, while the unstable parallel AA configuration decays into the energetically more favorable AB or BA ones sketched in Fig. 1. Here we start by comparing domains in bilayers with parallel AA, AB, and BA stackings. In the AB (BA) domain, the N anion (B cation) in the upper layer is over the hexagon center in the bottom layer, while the B (N) in the upper layer is right over the N (B) atom in the lower one. This configuration breaks inversion symmetry and gives rise to a vertical upward (downward) electric polarization of $1.46 \times 10^{-12}$ C/m by interlayer charge transfer (Fig. S1). The AB and BA domains can be morphed by the interlayer sliding along *a*, *b*, and *c* axis, as illustrated in Fig. 1. The intermediate saddle-point (SP) in the sliding pathway exhibits a purely in-plane polarization ($1.38 \times 10^{-12}$ C/m) as illustrated in Fig. S1. The out-of-plane polarization and energy barrier of the translation pathway is shown in Fig. 1b. We trained a deep potential (DP) model of *h*-BN using a machine learning method based on DFT datasets (for details, see Supplementary Materials [22]). The DP-calculated energies, forces, kinetic barriers, and phonon dispersion relations in Figs. S2, S3 [22], and Fig. 1b agree remarkably well with those computed directly by DFT, which implies that the DP model predicts structural and dynamic properties of arbitrary textures such as DWs, Moiré structures, and ferroelectric switching barriers with DFT accuracy.

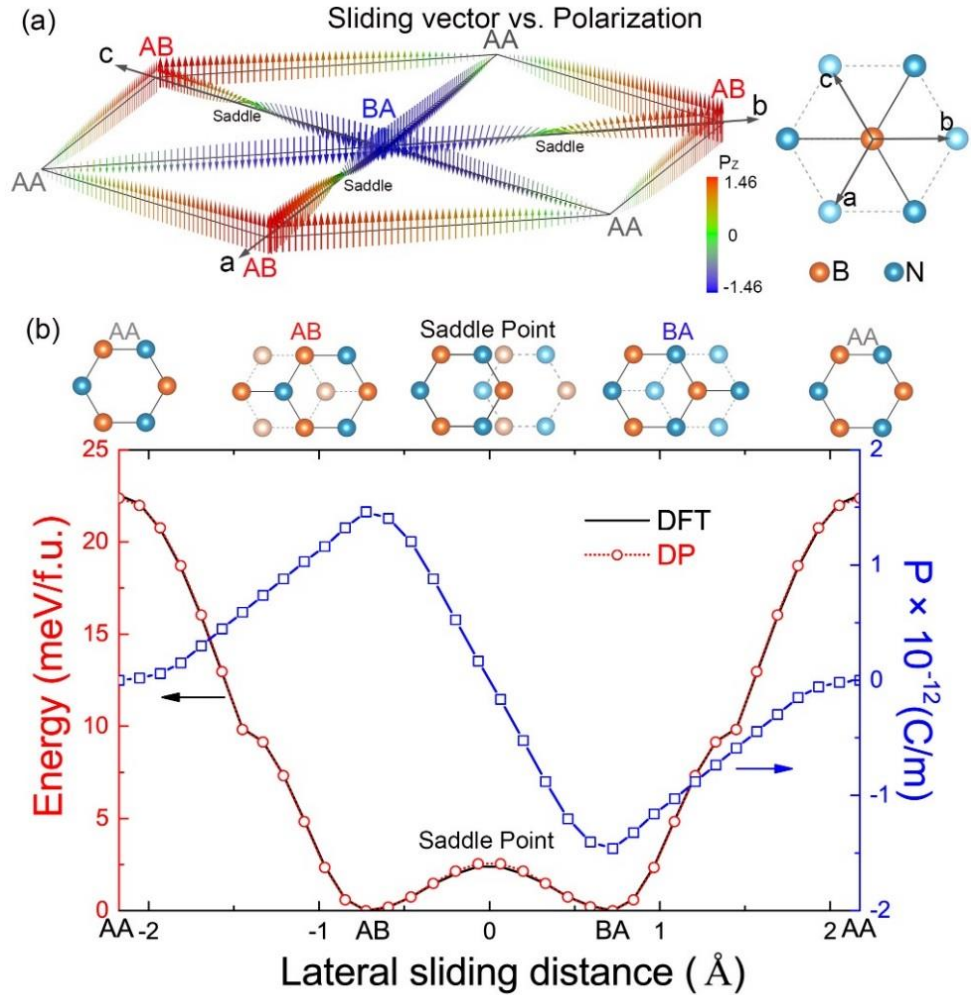

Fig.1: Ferroelectricity in h-BN bilayers as calculated by density functional theory (DFT) and the machine-learning potential model. (a) The electric polarization vector as a function of lateral sliding of the upper relative to the lower layer over a unit cell (right panel). The colors indicate the out-of-plane polarization $P_z$. (b) $P_z$ and total energy change along the translation pathway with minimum barrier. The AB and BA stacking corresponds to two opposite spontaneous polarization states that are separated by a saddle point, while the fully eclipsed AA stacking configuration has the highest energy.

Next, we explore the thermal stability of the ferroelectric polarization in freestanding $h$-BN bilayers from 50 K to 700 K by Deep Potential molecular dynamics (DPMD) simulations. According to Fig. S4a, the coherent monodomain remains stable at high temperatures because the stiff in-plane chemical bonds are nearly immune to thermal fluctuations [22]. As temperature increases from 0 K to 700 K, we observe out-of-plane ripples of the bilayer and an increase of the average interlayer distance, leading to a slight reduction of the polarization as shown in Fig. S4 [22]. For instance, the polarization decreases from $1.46 \times 10^{-12}$ C/m to $1.14 \times 10^{-12}$ C/m at 700 K, which should not seriously affect electronic devices applications. This robustness of the ferroelectric order confirms a recent theoretical prediction of a very high Curie temperature [6].

***Atomistic and polarization properties of DWs.*** −A DW separating AB and BA domains in *h*-BN bilayers is a continuous variation of the sliding vector ***b*** along one of the three-equivalent *a*, *b*, and *c* orientations in Fig. 1a that leads to a phase slip of π between domains of opposite polarization. We choose coordinates such that the DWs always lie in the *y* direction. They can be categorized into four types by the angle *φ* between DW and unit crystal vector ***b***, viz. 0°, 30°, 60°, and 90° as sketched by red arrows in Fig. 2 and Fig. S6 [22]. We construct periodic *h*-BN freestanding bilayer supercells (~30000 atoms) containing two DWs that are separated by 120 nm to prevent coupling. The DP model for the structural optimization leads to the supercells with DWs shown in Fig. S5 [22]. Fig. 2 and Fig. S6 [22] show zoom-in's on the DW region and the local atomic structure. In the 0° DWs the sliding vector ***b*** varies smoothly across the DW and the out-of-plane geometry remains flat (Fig. 2a). In contrast, we observe a huge buckling for three other DWs (Fig. 2b and Fig. S6 [22]) that may be associated with the flexoelectric effects in van der Waals free-standing films. Table SI shows that the 0° DW is most stable and DW energies increase linearly with the angle *φ* [22]. When stuck to a substrate that suppresses the buckling, the 30°, 60°, and 90° DW energies would shift to much higher energies.

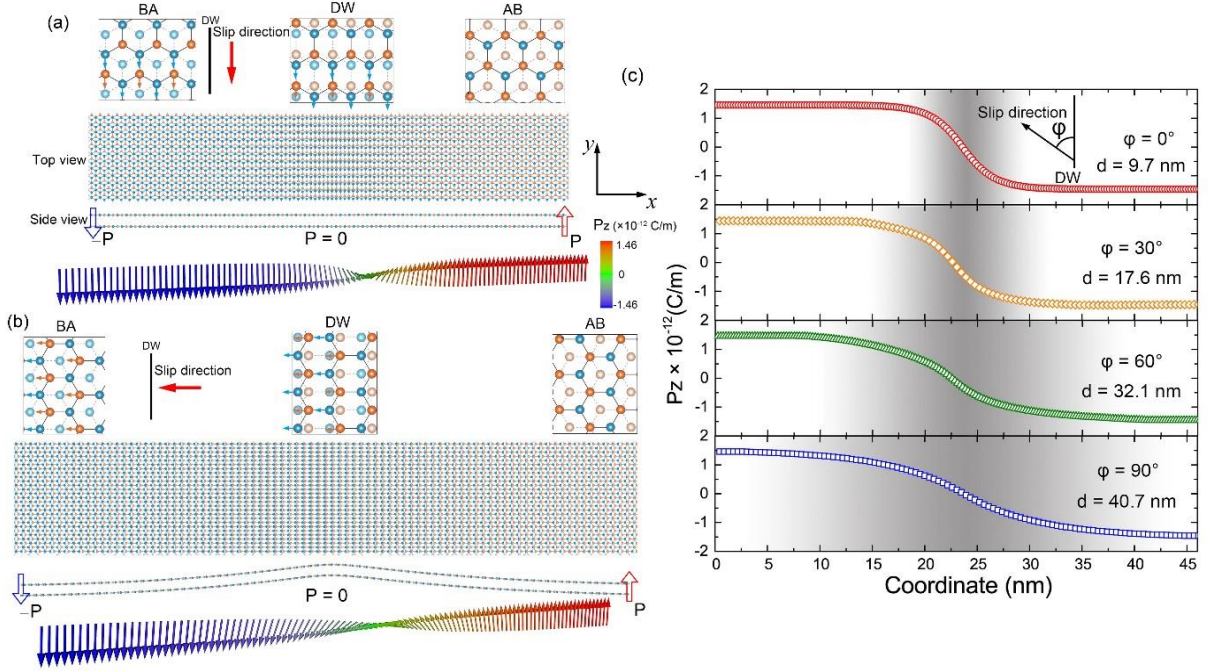

Fig. 2: Top and side views of atomic structure and polarization of DWs, which are a zoom-in of the large supercells in Figure S5. (a) 0° and (b) 90° DWs. (c) Out-of-plane polarization ($P_z$) across the DWs. The gray shading indicates the local polarization gradient.

A rotation of the polarization in conventional bulk ferroelectrics is energetically expensive because of the associated substantial elastic strains such that DW are usually of the "Ising type", i.e. collinear with a vanishing polarization at the DW center [11]. The DWs plotted in Fig. 2a, b exhibit textures closer to those of DWs in ferromagnets that are characterized by a rotation with fixed modulus. Here the polarization rotates with nearly constant $P$ ($\Delta P \backslash P \sim 5\%$). While the 0° DW has a Bloch configuration while the 90° DW exhibits Néel-like texture. The variation of the sliding vector $\boldsymbol{b}$ modulates the local out-of-plane polarization ($P_z$) as plotted in Fig. 2c for four different directions. The characteristic DW widths are 9.7 nm ($\varphi=0°$), 17.6 nm ($\varphi=30°$), 32.1 nm ($\varphi=60°$), and 40.7 nm ($\varphi=90°$), respectively, much larger than those in perovskite ferroelectrics (~0.7 nm). The proportionality of the DW width $w$ with angle $\varphi$ follows from a one-dimensional (1D) elastic model for the DW energy:

$$E = \int \left[\frac{\lambda_{1D}}{2}\left(\frac{\partial u_s}{\partial x}\right)^2 + V(u_s)\right] dx, \qquad (1)$$

where $u_s$ is the displacements of the top and the bottom layer along $x$, $\lambda_{1D}$ the corresponding 1D Lamé force coefficient, and $V(u_s)$ the interlayer potential energy density as a function of $u_s$. The width

$$w = \frac{u_0}{2}\sqrt{\frac{\lambda_{1D}}{\Delta}}, \qquad (2)$$

where $\Delta$ is the barrier height per unit length, minimizes the energy of a hyperbolic tangent shape (see Supplemental Material [22]). The DW widths in stacked ferroelectrics is so large because of both the ultralow ferroelectric switching barrier ($\Delta$) and the large in-plane monolayer stiffness ($\lambda_{1D}$). Since $\Delta$ does not depend on $\varphi$, the elastic constant $\lambda_{1D}$ governs $w(\varphi)$ (see Fig. S7 [22]). When fitting the 1D model to the DFT calculated distortion energies, $w$=10 nm ($\varphi$=0°), 14 nm ($\varphi$=30°), 30 nm ($\varphi$=60°), and 43 nm ($\varphi$=90°), consistent with the predictions of the full DP model (see Table SII [22]). While Eq. (2) predicts as well the width of ferromagnetic DWs when $\lambda_{1D}$ parameterizes the exchange interactions and $\Delta$ the magnetic anisotropy, $\lambda_{1D}$ strongly depends on direction, while the exchange interaction is isotropic. The phenomenology of DWs in stacking ferroelectrics with us therefore richer than that of simple ferromagnets.

***Dynamics of the DW motion.*** −In multidomain bilayer *h*-BN, vertical electric fields ($E_v$) or lateral shear stresses ($F_s$) can reverse the ferroelectric order. $E_v$ and $F_s$ are related by $F_s = E_v Z_{13}^*$, where $Z_{13}^*$ is a non-diagonal element of the Born effective charge tensor. For convenience, we regard the neighboring B and N ions in the top (bottom) layer as a single atom. A Berry phase calculation gives $Z_{13}^{top*} = 0.027$ and $Z_{13}^{bottom*} = -0.027$, which is two orders of magnitude smaller than in BaTiO$_3$. In the DPMD simulations, we can study the effects of additional strains $\pm F_s$ on top and bottom layer. The critical $F_s$ ($E_v$) at which the ferroelectric order of a single domain switches is $3.8 \times 10^{-3}$ eV/Å (~1.41 V/nm) at 100 K, and $3.5 \times 10^{-3}$ eV/Å (~1.32 V/nm) at 300 K. The critical $F_s$ is not sensitive to thermal fluctuation, in agreement with the predicted large thermal stability [6]. However, it is an order of magnitude larger than has been observed [3,4].

Many experiments report that the polarization switching by an electric field is mediated by DW motion [4]. When applying a small $E_v$ of 0.18 V/nm (or $F_s$ of $5 \times 10^{-4}$ eV/Å ) at 300K, we observe soliton-like motion of the DWs. Two 0° DWs with opposite polarity move closer to each other since the field decreases the energy density of the BA domain. After 9 ps they annihilate into single BA domain as shown in Fig. 3a (the motion of 90° DW is shown in Fig. S8 [22]). Fig. 3b shows time-dependent results for the BA domain size as a function of $E_v$. When $F_s < 7 \times 10^{-5}$ eV/Å, the DWs are immobile and the BA domain fraction remains constant, indicating an intrinsic pinning potential in the absence of disorder. For larger $F_s$, the BA domain growths linearly with time indicating a constant DW velocity of 6000 m/s at 0.18 V/nm (see Fig. S9 [22]), corresponding to switching times of ~15 ps for 100-nm-diameter electronic devices, thereby enabling the ultrafast non-volatile memory devices with low operation voltage. The critical $F_s$ for 0° and 90° DWs motion are $7 \times 10^{-5}$ eV/Å and $3 \times 10^{-4}$ eV/Å (corresponding to $E_v$=0.026 V/nm and 0.11 V/nm), recalling that the monodomain switching field (force) of $3.8 \times 10^{-3}$ eV/Å (1.41 V/nm) is larger by two orders of magnitude.

Next, we address ferroelectric skyrmions under electric fields at room temperature. Fig. S10 [22] is a top view of the domain pattern and polarization distribution of an isolated skyrmion bubble under a perpendicular electric field of ~5 V/nm in the 4-nm-diameter circle. The ferroelectric is here a freestanding bilayer with AB domain. The topological charge of ferroelectric skyrmions is an integer that represents the bit states "1" and "0". Magnetic skyrmions usually require stabilization by a local external magnetic field, complicating their integration into nanoscale electronic devices. No such issues hamper the control of ferroelectric skyrmion by electric fields, which facilitates ultra-dense information storage devices. Our results open a rich parameter space to generate and manipulate skyrmions or other topological defects in stacking-engineered ferroelectrics.

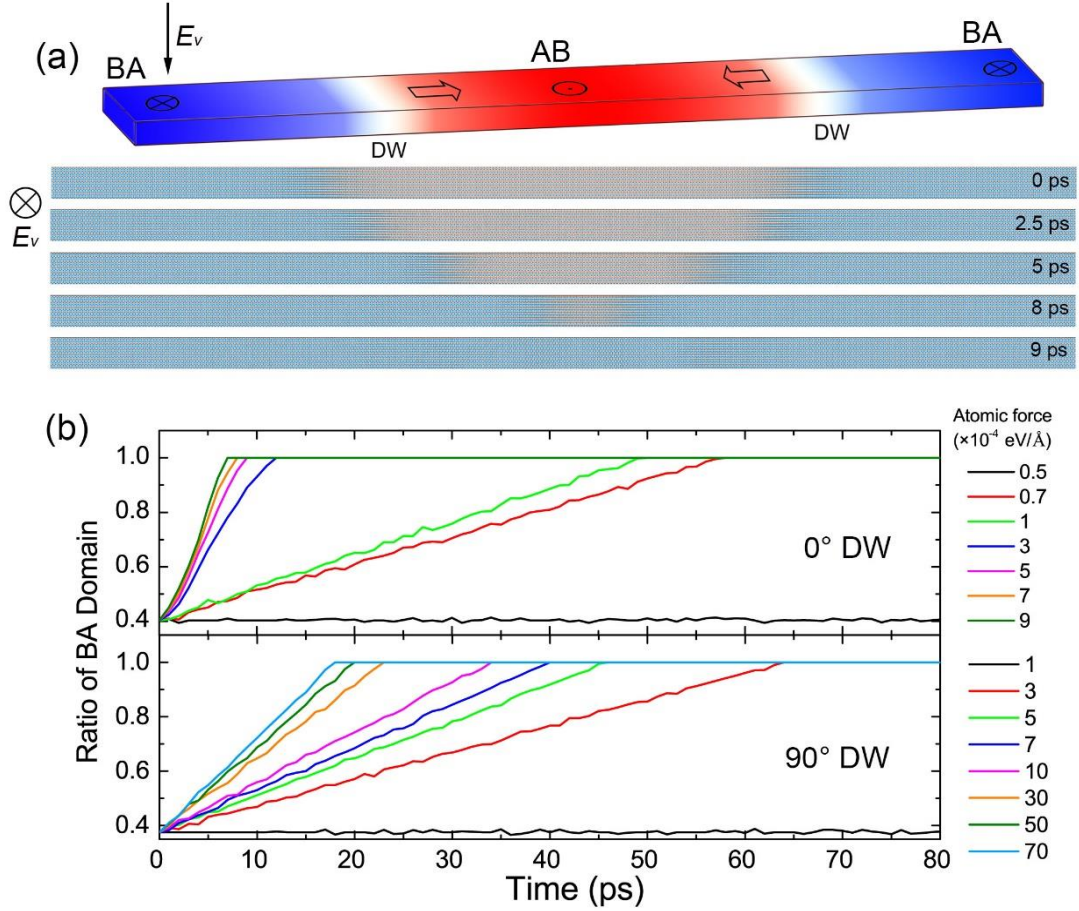

Fig. 3: Time-dependent simulations of DW motion in h-BN bilayers at 300 K. (a) Ferroelectric textures with two 0° DWs under an external lateral shear force $F_s$ (i.e. perpendicular $E_v$) of $5\times10^{-3}$ eV/Å switched on at $t=0$. The snapshots at 0, 2.5, 5, 8, and 9 ps reveal fast motion. The DWs annihilate each other after ~9 ps. (b) Evolution of the AB domain with time under shear stresses for 0° and 90° DWs.

*Super-paraelectricity in Moiré structure.* −Twisting two van der Waals layers by a small angle gives rise to Moiré patterns that according to several experiments cause a ferroelectric response [3,4,9,26]. On the other hand, recent analytical models do not find a macroscopic polarization of a commensurate Moiré structure at zero field [27-29]. We clarify this contradiction by carrying out a large-scale atomistic simulation of a Moiré structures of *h*-BN bilayers with a small 0.385° twist, which corresponds to a lateral supercell containing 355012 atoms as displayed in Fig. S11a [22]. Initially, the areas of AB, BA, and AA domains are the same. Allowing atoms to relax in the DP model, the AB, BA domain expand into triangles, while the high-energy AA domain shrinks to a point as shown in Fig. S11b [22]. The 0° DWs accommodate the conserved global twist when the Moiré structure disintegrates into commensurate AB and BA domains (see Fig. S11c [22]). In the corners of the triangular domains, fully eclipsed AA domains persist without twist. A periodic triangular configuration agrees well with observed patterns [3,4,9,26]. The fully relaxed configuration is very stable against thermal fluctuations as shown in Fig. S12 [22].

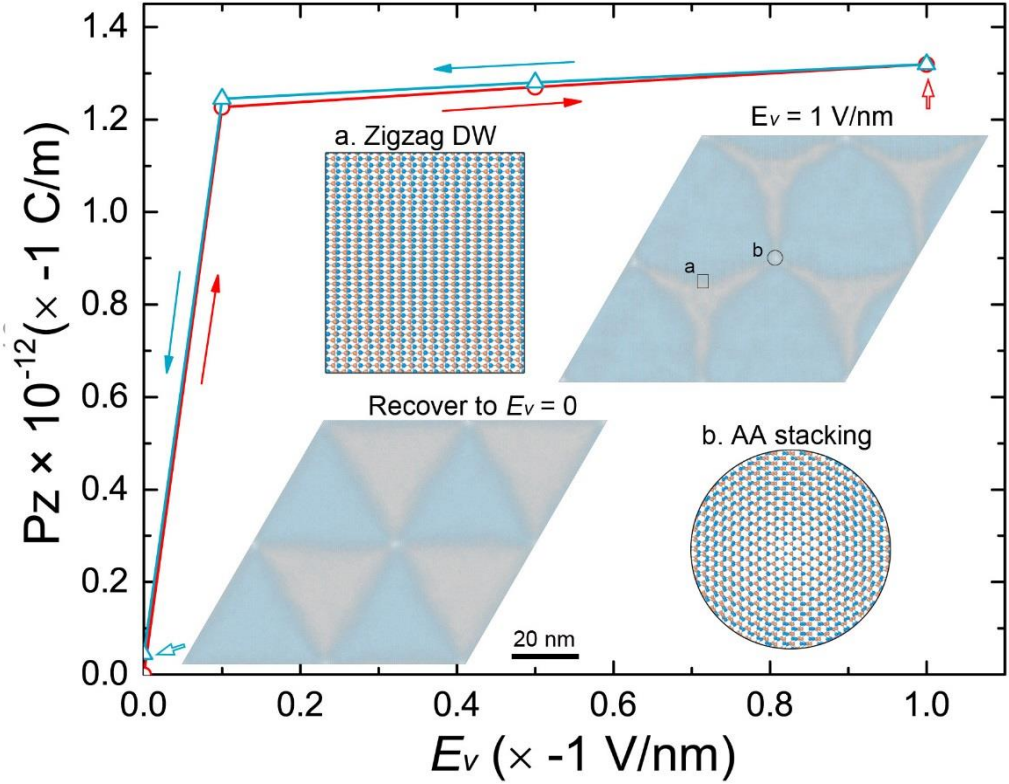

Fig. 4: Spatially averaged of polarization of a twisted Moiré bilayer BN as a function of a perpendicular electric field ($E_v$). The insets show the nearly reversible polarization texture for the largest and zero $E_v$.

We study the response of the Moiré structure and polarization to a vertical electric field by DP models trained by the DFT calculations for $E_v$ = 0.1, 0.5, and 1 V/nm (Supplementary Materials S1 [22]). For $E_v$ = 1 V/nm, the BA domain changes from a triangular to a hexagonal-like pattern, while the AB domain shrinks (see right upper insets of Fig.4), leading to a net macroscopic out-of-plane polarization $P_z = 1.32 \times 10^{-12}$ C/m. The main plot in Fig. 4 shows the spatial average of $P_z$ as a function of $E_v$ (P-E hysteresis). We attribute the initial increase and rapid saturation at an ultra-low critical

$E_v$ = 0.026 V/nm for 0° DWs motion in terms of pinning by the AA domain that prevents an increase of $P_z$ for larger $E_v$. The polarization recovers its original texture when removing the field again, as shown in left bottom inset of Fig.4. At zero field AB and BA domains have the same size and the macroscopic polarization nearly vanishes. The absence of a hysteresis in the *P-E* diagram of the twisted configuration indicates super-paraelectric rather than ferroelectric order [27]. We conclude that the observed ferroelectric response of Moiré structures is caused by defects that pin the domain wall motion that would equalizes the AB and BA domain sizes such that a net polarization persists.

In summary, using a machine-learning-based Deep-Potential model, we predict rich polarization profiles in *h*-BN bilayers, the archetypal van der Waals stacking ferroelectric. We discuss four types of ferroelectric DWs and topological electric skyrmions. We reveal that DWs lower the critical electric switching field by two orders of magnitude and reduce switching times to picoseconds. Moiré structures in twisted bilayers can be tuned reversibly by a vertical electric field and behave more like a super-paraelectric than a ferroelectric. Our simulations reveal the importance of domain walls in understanding dynamic properties of ferroelectricity and provide necessary theoretical guidance to the future exploration of van der Waals ferroelectrics and its applications.

This work was supported by the National Key R&D Program of China (Grants No. 2021YFA0718900 and No. 2022YFA1403000), the Key Research Program of Frontier Sciences of CAS (Grant No. ZDBS-LY-SLH008), the National Nature Science Foundation of China (Grants No. 11974365 and No. 12204496), the K.C. Wong Education Foundation (Grant No. GJTD-2020-11), and the Science Center of the National Science Foundation of China (Grant No. 52088101). GB and PT were supported by JSPS Kakenhi Grants No. 19H00645 and 22H04965.

# Supplementary Materials

# Ultrafast switching dynamics of the ferroelectric order in stacking-engineered ferroelectrics

Ri He[1#], Bingwen Zhang[1,2#], Hua Wang[3], Lei Li[4], Tang Ping[5], Gerrit Bauer[5,6], Zhicheng Zhong[1,7*]

[1]Key Laboratory of Magnetic Materials Devices & Zhejiang Province Key Laboratory of Magnetic Materials and Application Technology, Ningbo Institute of Materials Technology and Engineering, Chinese Academy of Sciences, Ningbo 315201, China

[2]Fujian Key Laboratory of Functional Marine Sensing Materials, Center for Advanced Marine Materials and Smart Sensors, College of Material and Chemical Engineering, Minjiang University, Fuzhou 350108, P. R. China

[3]School of Micro-Nano Electronics, Hangzhou Global Scientific and Technological Innovation Center, Zhejiang University, Hangzhou 310027, China

[4]Frontiers Science Center for Flexible Electronics, Xi'an Institute of Flexible Electronics (IFE) and Xi'an Institute of Biomedical Materials & Engineering, Northwestern Polytechnical University, 127 West Youyi Road, Xi'an 710072, China

[5]WPI-AIMR, Tohoku University, 2-1-1 Katahira, Sendai 980-8577, Japan

[6]Kavli Institute for Theoretical Sciences, University of the Chinese Academy of Sciences, Beijing 10090, China

[7]China Center of Materials Science and Optoelectronics Engineering, University of Chinese Academy of Sciences, Beijing 100049, China

---

[#] These authors contribute equally to this work.

[*] zhong@nimte.ac.cn

# S1. Calculation Methods

a. **The construction of a Deep Potential model**

Machine-learning-based Deep Potential (DP) model is an emerging method of descriptions of atomic interactions that fill the gap between accurate first principles DFT calculations and efficient large-scale atomistic simulations. Carefully choosing the *h*-BN configurations for the training dataset is crucial for an accurate DP model. Here we use the DP Generator (DP-Gen) to generate training data that efficiently cover a most of the relevant configurational space [1], which is a concurrent learning strategy. The workflow of each iteration includes three main steps: (1) Training the deep learning potentials, (2) exploring configurations by deep potential molecular dynamics (DPMD) simulations, and (3) labeling configurations according to certain criteria and adding them to the training dataset.

We start with AA, AA', AB, BA, AB' and BA' bilayer configurations and their intermediate transition states that have been optimized by density functional theory (DFT) calculations. In the first iteration of the DP-Gen workflow, the initial training dataset contains 1000 randomly perturbed 3×3 supercells containing 36 atoms by displacement of the atomic coordinates from the ground states by <0.01 Å that cause in-plane strains of −0.003 to 0.003. Next, we train four different DPs by the DPMD method [2] with different values of deep neural network parameters. In the exploration step one of the DPs is used for molecular dynamics (MD) simulations for different pressures (1 to 50000 bar) and temperatures (10 to 1200K) to explore the configuration space. The other three DPs predict the atomic forces for all configurations visited by the MD trajectories. The maximum deviation of the forces $F_i$ on an atom *i* from the average $\langle F_i \rangle$ as predicted by the four DPs

$$\sigma_f^{max} = \max\sqrt{\langle |F_i - \langle F_i \rangle|^2 \rangle},$$

can be used as a criterion to label different configurations. $\sigma_f^{max} < \sigma_{low}$ indicates the accuracy of the current DP as set by $\sigma_{low}$. A configuration with $\sigma_f^{max} > \sigma_{high}$ is highly distorted and considered as a failure with a threshold set by $\sigma_{high}$. The failure ratio is often high in the first several iterations. Only configurations satisfying $\sigma_{low} < \sigma_f^{max} < \sigma_{high}$ are selected as candidates for subsequent self-consistent DFT calculations and are added to the training dataset for the next iteration. $\sigma_f^{max}$ can be also used as the convergence criterion for DP-Gen iterations that are considered converged when the accuracy ratio is larger than 99.5 %. Here, $\sigma_{low}$ and $\sigma_{high}$ are set to 0.05 and 0.15 eV/Å. Iterating the above procedure 23 times generates ~10 000 training configurations. For more details of the DP-Gen process, please refer to the original literature [1] or our previous work on $SrTiO_3$ and $ZrW_2O_8$ models [3,4].

Next, we train the DP model by the DeePMD-kit code [5] in which the total potential energy of a configuration is assumed to be a sum of atomic energies mapped from a descriptor through an embedding network. The descriptor characterizes the local environment of an atom within a cutoff radius set here to $R_c$ = 6 Å that includes a maximum number of 200 B and N atoms. The translational, rotational, and permutational symmetry of the descriptor are preserved by another embedding network while smoothing the discontinuities introduced by the fixed $R_c$. The sizes of the embedding and fitting networks are (25, 50, 100) and (240, 240, 240), respectively. The loss function have the same form as in our recent work [3]. The weight coefficients of the energy, atomic force and virial terms in the loss functions change during the optimization process from 0.2 to 1, 500 to 1, and 0.02 to 0.2, respectively. The DPs are trained with 1.800.000 steps with learning rates exponentially decaying by $10^{-3}$ to $3.5 \times 10^{-8}$. A model compression scheme was applied to boost the computational efficiency [6].

**b. DFT calculations**

We use the Vienna ab initio simulation package (VASP) [8] with projector augmented wave method and the generalized gradient approximation (GGA) with Perdew-Burke-Ernzerhof of exchange-correlation functional [7]. We adopt a plane-wave cutoff energy of 650 eV in the structural relaxations calculations. A large distance of $c >$ 15 Å along the out-of-plane direction eliminates interlayer interactions. A dispersion correction with the optB86b functional in structure related calculations [7] improves the accuracy of structural properties of layered materials. The optimized lattice constant of AB stacked bilayer h-BN is $a = b = 2.510$ Å with interlayer distance 3.256 Å. The Berry phase method leads to a calculated perpendicular polarization of the AB domain of $P_z=1.46 \times 10^{-12}$ C/m, which is somewhat smaller than the $2.08 \times 10^{-12}$ C/m predicted by previous theoretical work [9], probably deriving from the different van der Waals functional. We study the Moiré structures and associated polarizations by self-consistent DFT calculations with different out-of-plane electric fields of 0.1 V/nm, 0.5 V/nm, and 1 V/nm that generate new training datasets and corresponding DP models. We define a domain wall energy

$$\gamma(N) = \frac{E_{DW}(N) - E_{mono-domain}(N)}{2S},$$

where $E_{DW}$ and $E_{mono-domain}$ denote the total energy of supercells with and without domain walls, $N$ is the number of unit cells in the supercell, $S$ is the area of the system. We also calculate phonon properties in h-BN bilayer using density-functional-perturbation theory (DFPT) [10].

### c. Molecular dynamics (MD) simulations

The MD simulations in the exploration step are carried out by the LAMMPS code with periodic boundary conditions [11]. We adopt the isobaric-isothermal (NPT) ensemble with temperature set from 10 to 1200 K and pressure set from 1 bar to 5 GPa, because the explorative temperature and pressure should exceed those in the simulations. A Nose-Hoover thermostat and Parrinello-Rahman barostat control temperature and pressure, respectively [12,13]. The time step in simulations is set to 1 fs. The DPMD simulations of 0°, 30°, 60°, and 90° DWs start with $2 \times \sqrt{3}$ cell building blocks (see Fig. S7), and $240 \times 20 \times 1$ and $300 \times 20 \times 1$ supercells. Fig. S9a displays the superlattice model for a Moiré bilayer with 0.385° twist angle that contains 355012 atoms.

## S2. Domain wall width

The elastic energy of a one-dimensional (x-direction) ferroelectric texture reads

$$E = \int [\frac{\lambda_{1D}}{2}(\frac{\partial u_s}{\partial x})^2 + V(u_s)]dx,$$

where $u_s$ is the displacements of the top and the bottom layer along $x$, $\lambda_{1D}$ the corresponding 1D Lamé force coefficient, and $V(u_s)$ the interlayer potential energy density as a function of $u_s$. Minimizing with respect to $u_s$ by

$$\delta E = \delta \int_{-\infty}^{+\infty} [\frac{\lambda_{1D}}{2}(\frac{\partial u_s}{\partial x})^2 + V(u_s)]dx = 0$$

leads to

$$\lambda_{1D}\frac{\partial^2 u_s}{\partial x^2} = \frac{\partial V(u_s)}{\partial u_s}$$

We model $V(u_s)$ by a double-barrier potential[14],

$$V(u_s) = \frac{\Delta}{u_0^4}(u_s^2 - u_0^2)^2$$

where $\Delta$ is the barrier height per unit area and the potential minima are at $\pm u_0$. The above equations yield:

$$\lambda_{1D}\frac{\partial^2 u_s}{\partial x^2} + \frac{4\Delta}{u_0^2}u_s - \frac{4\Delta}{u_0^4}u_s^3 = 0$$

with trivial solution $u_s=0$ and DW solution

$$u_s(x) = \pm u_0 \tanh\frac{2(x - x_0)}{w}$$

where $x_0$ denotes the central position of the DW and

$$w = \sqrt{\frac{2\lambda_{1D}}{V''(u_0)}} = \frac{u_0}{2}\sqrt{\frac{\lambda_{1D}}{\Delta}}$$

is the characteristic width.

## S3. The Lamé parameters λ of stacking DWs

We estimate the force constant $\lambda_{1D}$ in the one-dimensional model by DFT calculations of $2 \times \sqrt{3}$ supercells. Fig. S7 a, b, c, d corresponds to 0°, 30°, 60°, and 90° DWs, respectively. The red arrow indicates the offset vector of the B atoms from their equilibrium positions with modulus 0.1 Å. The calculated energies of the four configurations in Fig. S7 are listed in Table 1. According to Eq (2) in the main text

$$(w_{\varphi_1})^2 - (w_{\varphi_2})^2 = C\Delta E_{12},$$

where $C$ is a proportionality constant and $\Delta E_{12}$ is the energy difference of supercells corresponding to the four types of DW in Fig. S7.

Table SI. The energies of 0°, 30°, 60°, and 90° DWs.

| Types of DWs | DW Energy (DP) eV/Å² |
|---|---|
| 0° | 21.97477 |
| 30° | 22.21938 |
| 60° | 22.62523 |
| 90° | 22.77754 |

Table SII. The characteristic widths of 0°, 30°, 60°, and 90° DWs.

| Type | Relative energy of the configurations in Fig. S7/meV | DW width (model fit) /nm | DW width (DP) /nm |
|---|---|---|---|
| 0° | 0 | 10 | 9.7 |
| 30° | 6.20 | 14 | 17.6 |
| 60° | 53.14 | 30 | 32.1 |
| 90° | 113.91 | 43 | 40.7 |

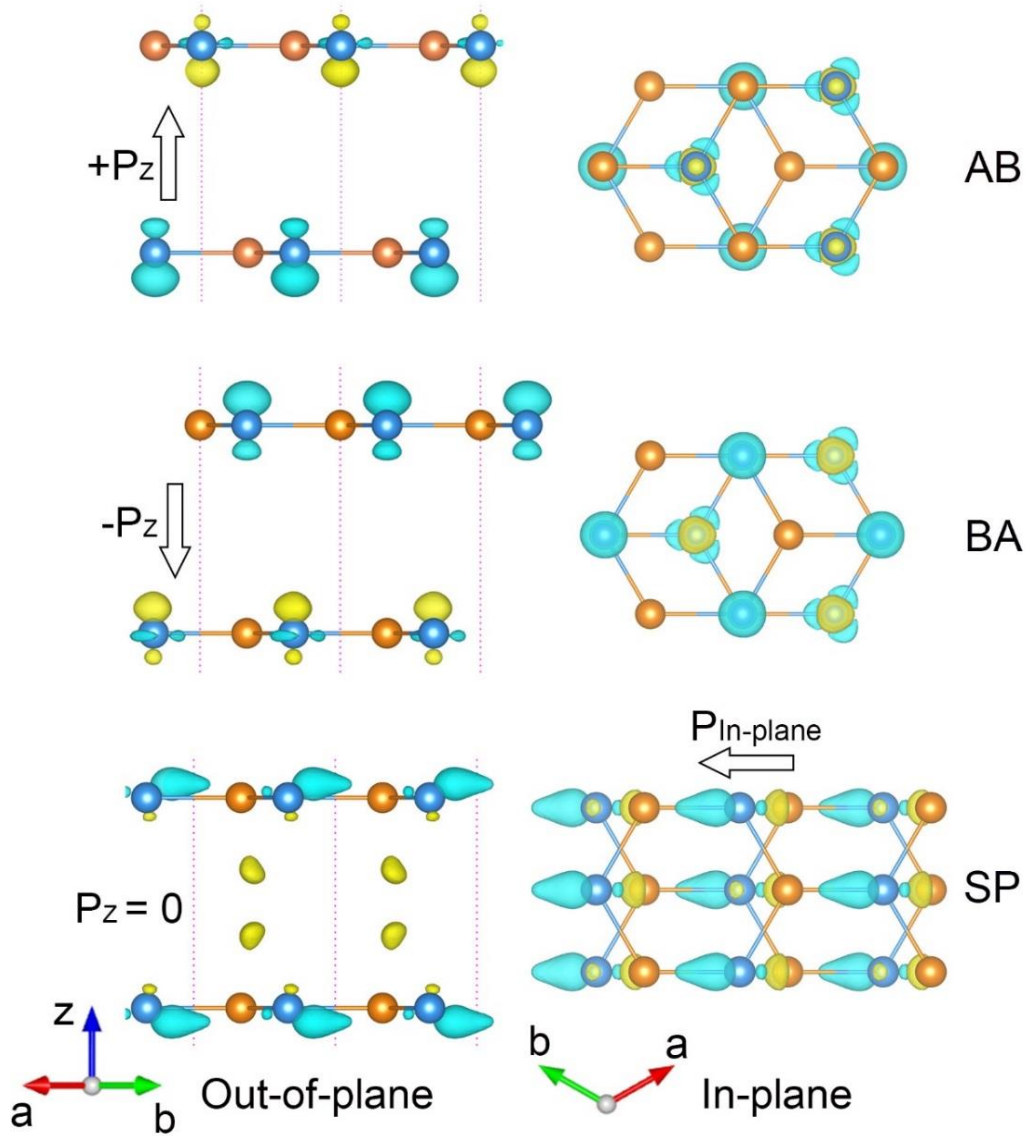

Fig. S1: The atomic arrangement and charge density difference for AB, BA, and SP stacking configurations of *h*-BN bilayer. Nitrogen and boron atoms are shown in yellow and blue, respectively. The distortions in the vertical alignment of nitrogen and boron atoms in the AB (BA) configurations induce charge transfer and upward (downward) electric polarizations. The result is a permanent and macroscopic out-of-plane electric dipole, while the SP stacking configuration creates an (unstable) in-plane electric dipole. The iso-surface level of the charge density difference maps is set to 0.00017 e/Å$^3$.

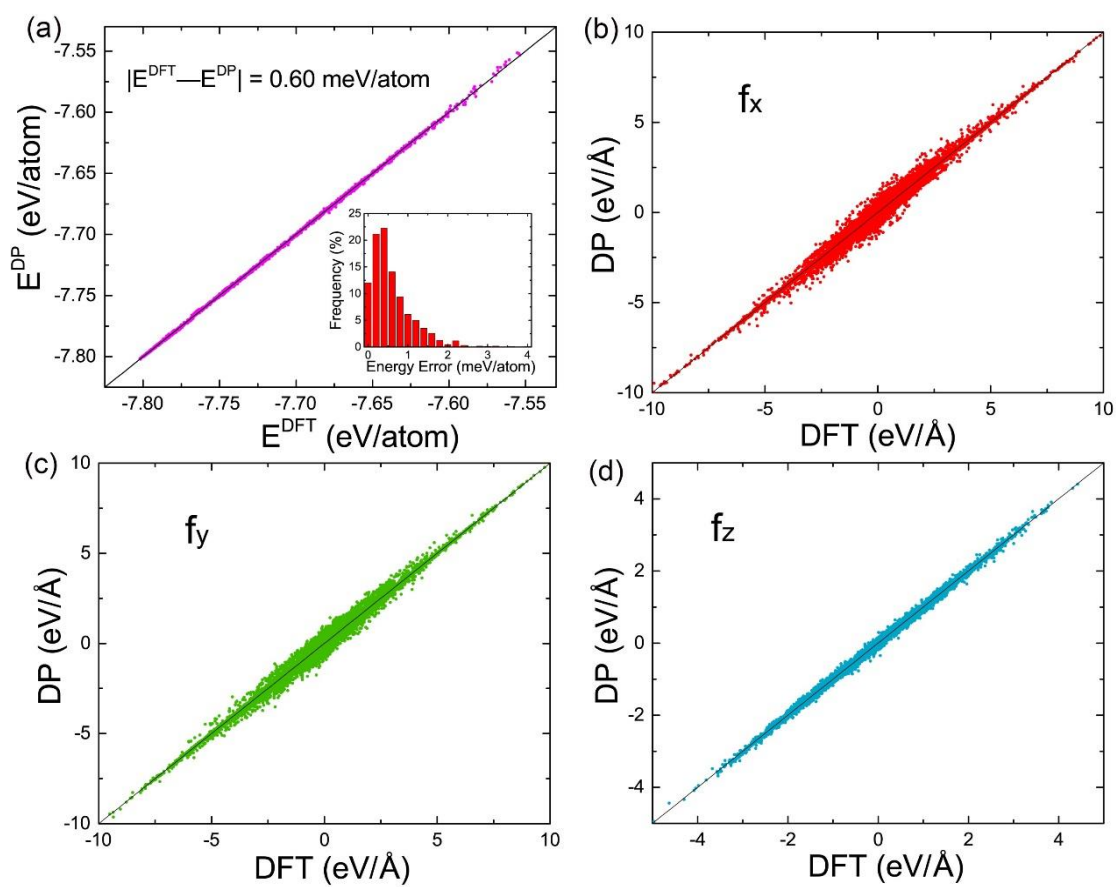

Fig. S2: The benchmark test of DP against DFT results. Comparison of energies (a) and atomic force components along $x$ (b), $y$ (c), $z$ (d) axis of the DP against DFT calculations for all configurations in the training dataset.

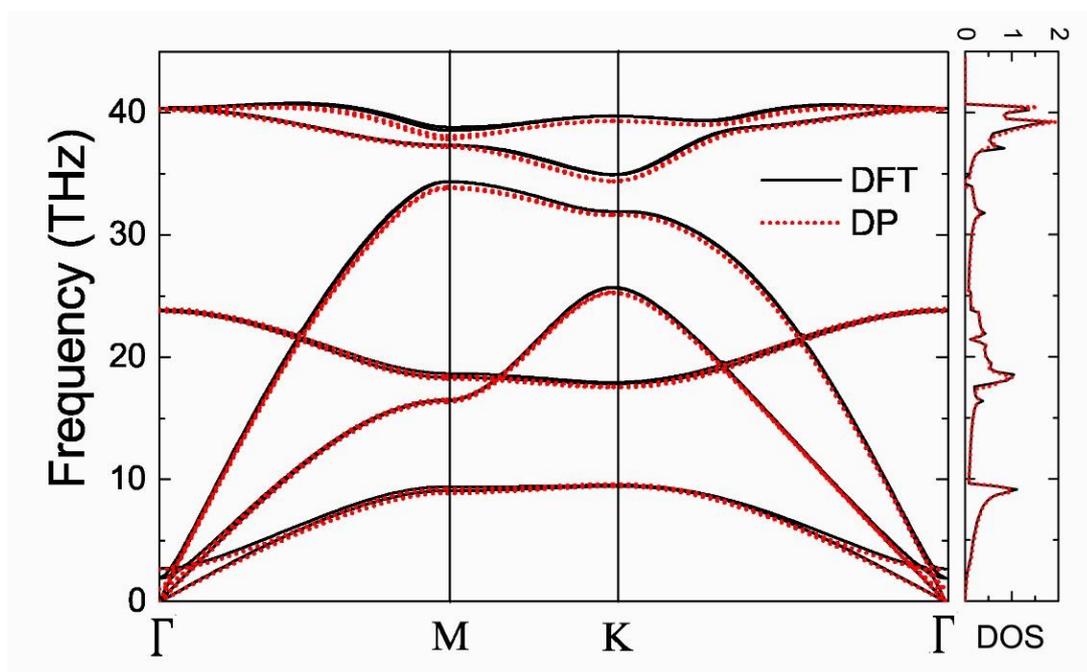

Fig. S3. The phonon dispersion relations and density of states (DOS) of the AB domain of *h*-BN bilayer calculated by the DP model and DFT.

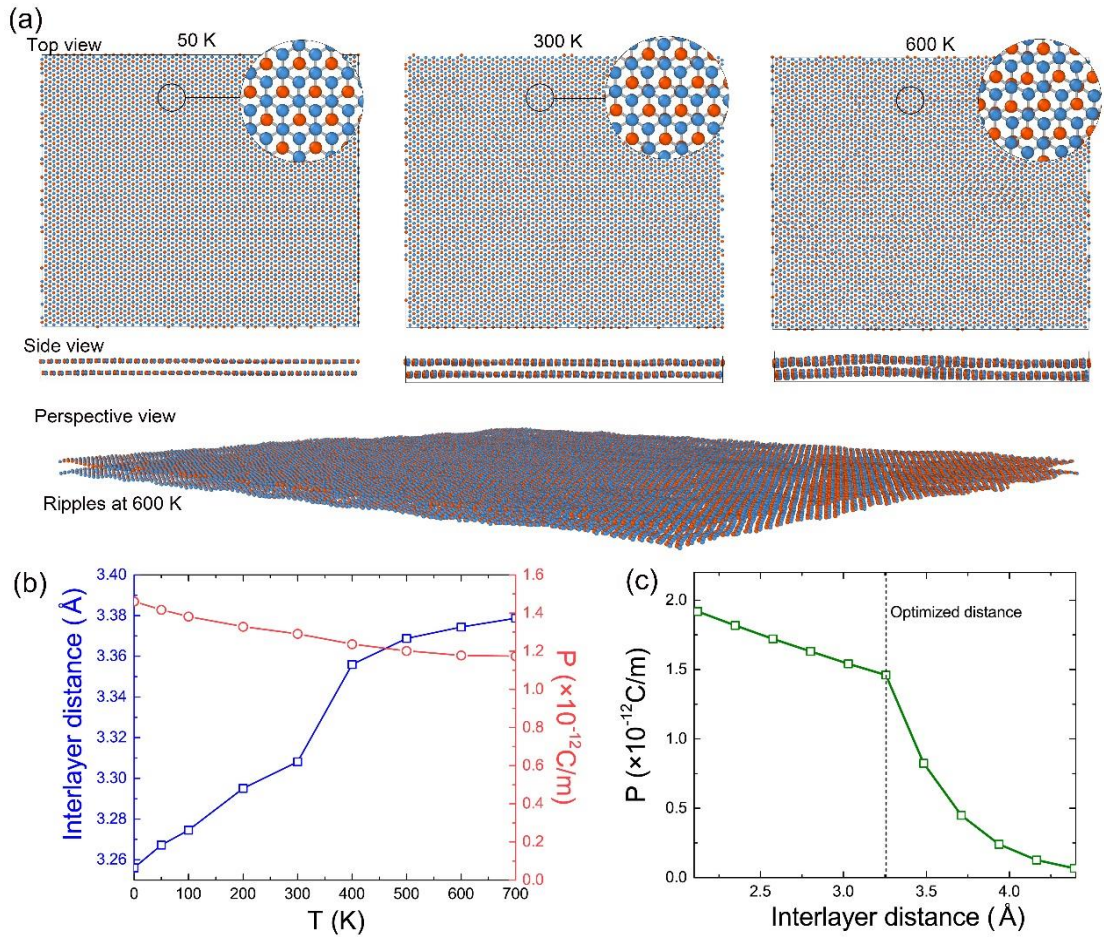

Fig. S4: AB stacking configuration of bilayer h-BN (40000 atoms supercell) at temperatures from 0 K to 700 K from the DPMD simulation. (a) Atomic structures. (b) The temperature-dependent interlayer distance and polarization. (c) The perpendicular polarization as a function of interlayer distance.

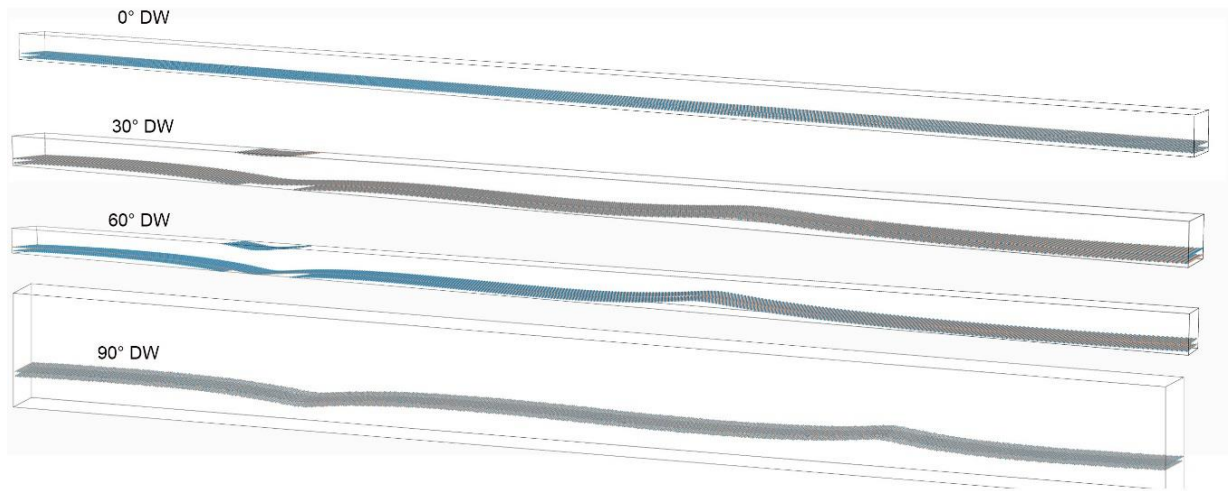

Fig. S5: 0˚, 30˚, 60˚, and 90˚ DW in bilayer h-BN with periodic supercells (including 30000-40000 atoms) optimized by the DP model.

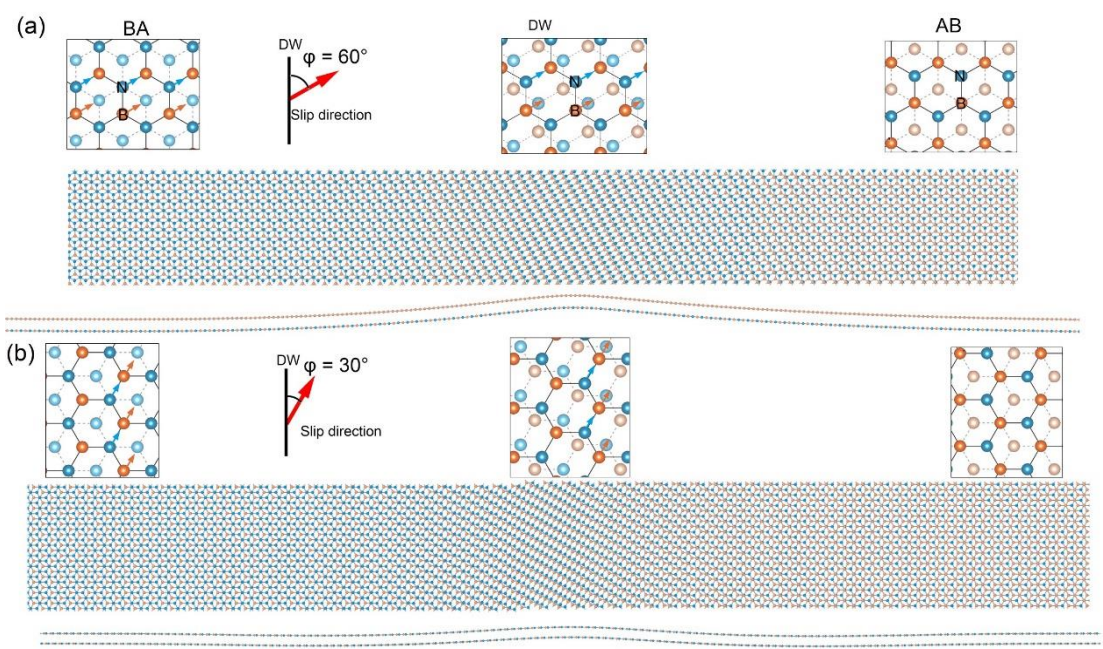

Fig. S6: Building blocks of BA and AB stacking domains geometry of supercells with 60° and 30° DWs. These pictures are zoom-in on the full supercells in Figure S5.

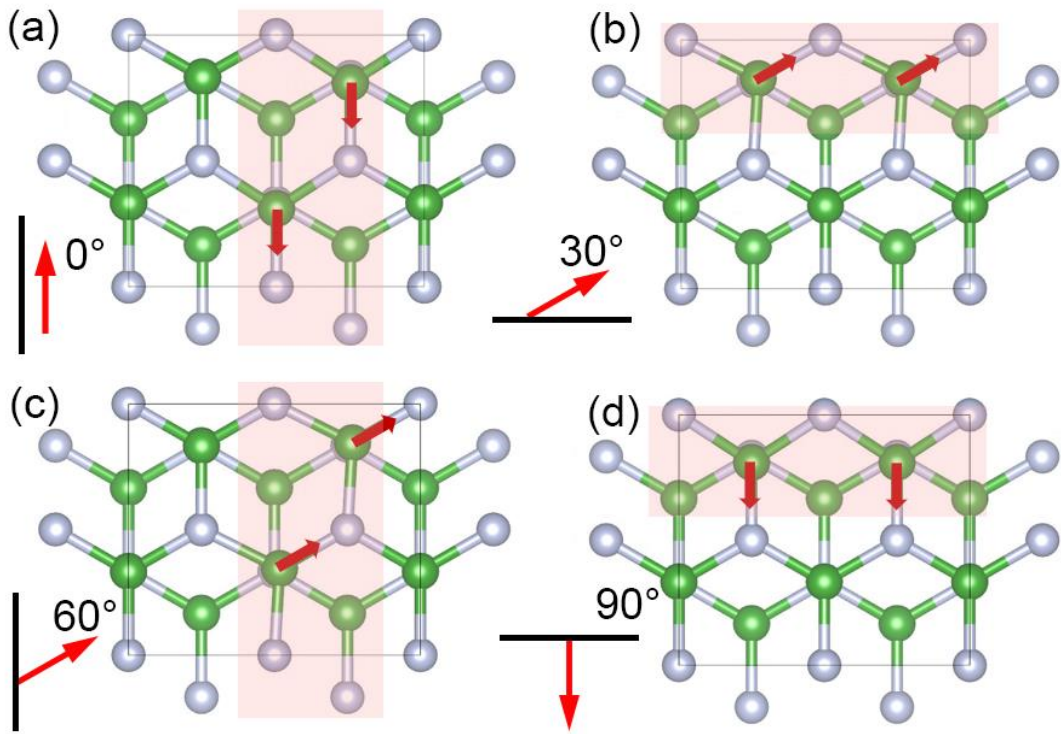

Fig. S7: The 2 × √3 supercells with different directions of the displacement by 0.1 Å of the B atoms from their equilibrium positions that corresponds to (a) 0°, (b) 30°, (c) 60°, and (d) 90° domain walls, respectively. The elastic energy cost determines the 1D Lamé coefficient ($\lambda_{1D}$) in Eq. (1).

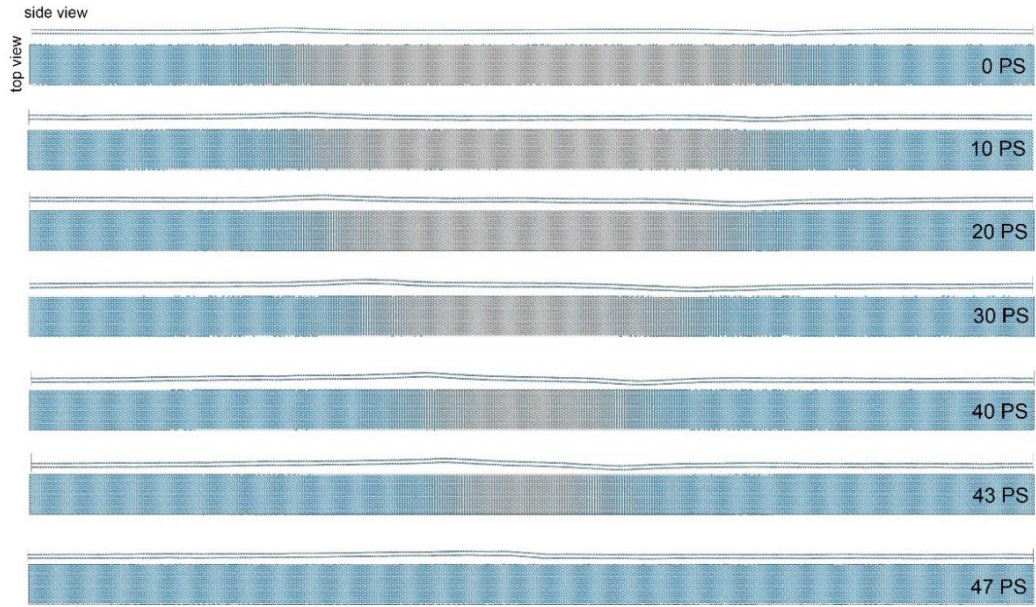

Fig. S8: Domain patterns that reveal the motion and annihilation of 90°DWs under a perpendicular electric field.

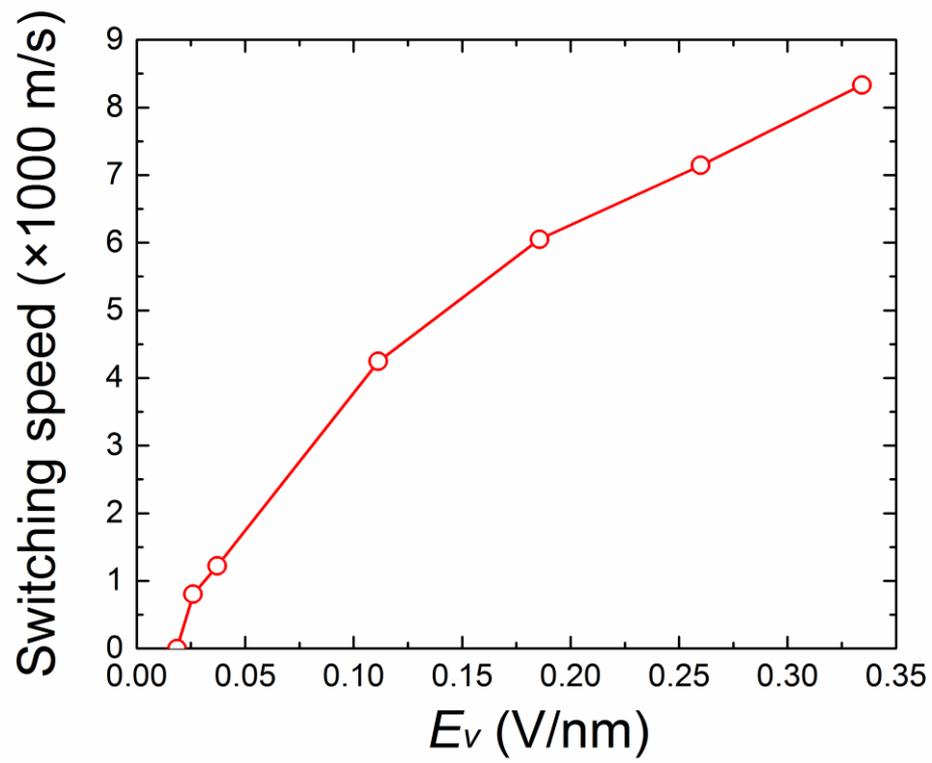

Fig. S9: Average domain switching speed as a function of applied perpendicular electric fields.

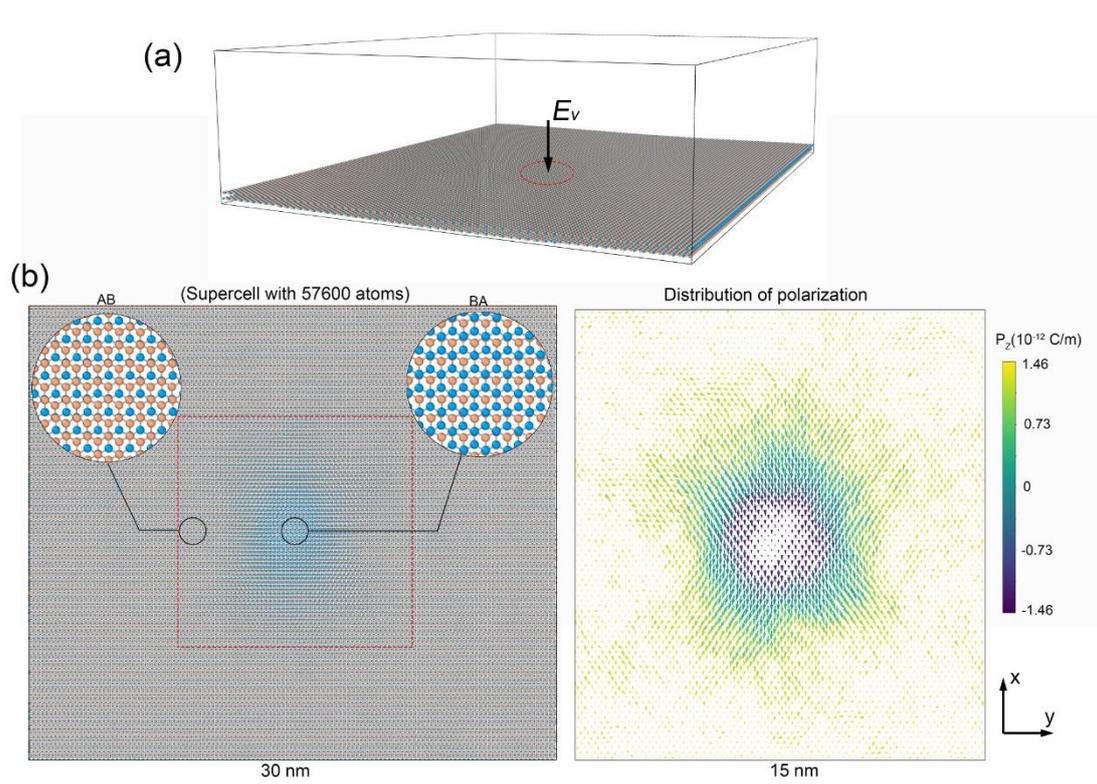

Fig. S10: Top view of a (b) ferroelectric domains pattern and corresponding polarization distribution of freestanding h-BN bilayers under a local perpendicular electric field as illustrated in (a).

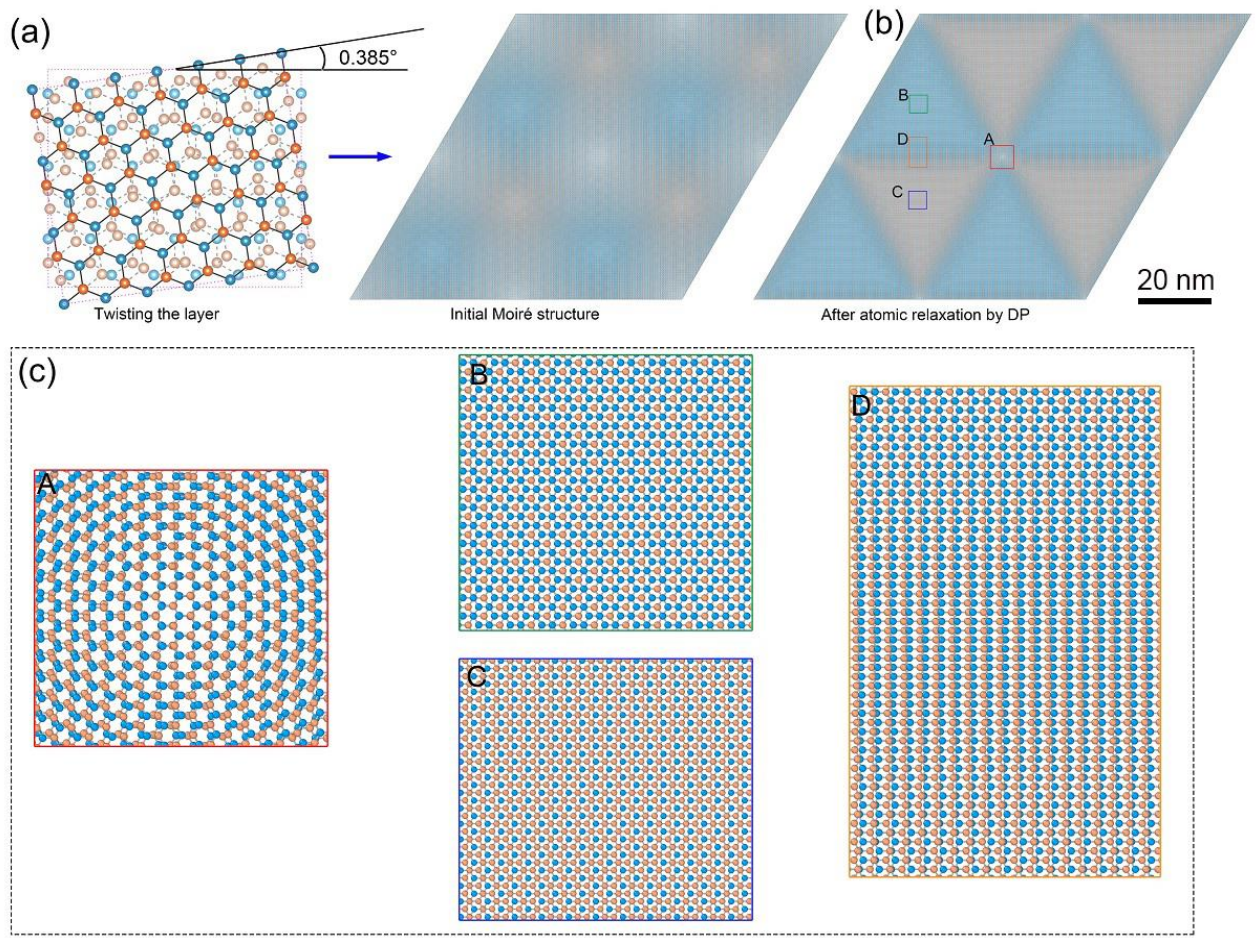

Fig. S11: (a) The initial Moiré structure corresponding to a 0.385° twisting angle between two h-BN layers containing 355012 atoms. The areas of AB, BA, and AA stacking domains are equal. (b) After atomic relaxation by DP model, the areas of the AB, BA domains expanded while that of the AA domain shrinks to a point. (c) The Moiré structure disintegrates into large domains of commensurate triangular AB and BA domains separated by 0° zigzag-type DWs that accommodate the global twist. Fully eclipsed AA domains without twist appear in each corner of the triangular domains.

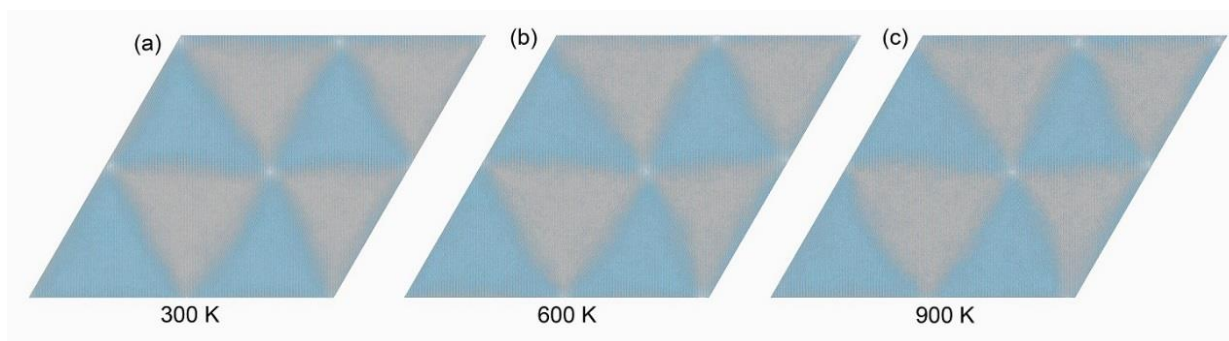

Fig. S12: The twisted bilayer superlattice at (a) 300 K, (b) 600 K, and (c) 900 K, indicating high thermodynamical stability of the triangular topology.